\documentclass[12pt]{iopart}
\usepackage{iopams}
\usepackage{graphicx}
\usepackage{caption}
\usepackage{color,ulem}

\begin{document}

\title{Thermal noise of gram-scale cantilever flexures}
\author{Thanh T-H. Nguyen$^1$, Bram J.~J. Slagmolen$^1$, Conor M. Mow-Lowry$^2$, John Miller$^1$, Adam Mullavey$^1$, Stefan Go\ss ler$^1$, Paul A. Altin$^1$, Daniel A. Shaddock$^1$, and David E. McClelland$^1$}

\address{$^1$Centre for Gravitational Physics, The Australian National University, Science Rd 38a, 0200 Canberra, ACT, Australia\linebreak $^2$School of Physics and Astronomy, The University of Birmingham, Edgbaston, Birmingham, B15 2TT, UK}

\ead{\mailto{thanhtruc.nguyen@anu.edu.au}}

\date{\today}

\begin{abstract}

We present measurements of thermal noise in niobium and aluminium flexures. Our measurements cover the audio frequency band from 10\,Hz to 10\,kHz, which is of particular relevance to ground-based interferometric gravitational wave detectors, and span up to an order of magnitude above and below the fundamental flexure resonances at 50\,Hz -- 300\,Hz. Our results are well-explained by a simple model in which both structural and thermoelastic loss play a role. The ability of such a model to explain this interplay is important for investigations of quantum-radiation-pressure noise and the standard quantum limit.

\end{abstract}

\pacs{}

\maketitle


\section{Introduction}
\label{sect:introduction}

Thermal fluctuations have become one of the fundamental sources of noise in high-precision experiments and are of increasing interest to many research groups \cite{Jiang2011b,Lubbe2013,Kessler2012,Cole2013,Purdy2013}. A prime example is interferometric gravitational wave detectors, in which the mitigation of thermal noise constitutes one of the most challenging aspects of the design of mirrors and suspension systems \cite{Agatsuma2010b, Cumming2014, Hammond2012}. It is also crucial to understand the role of thermal noise in experiments investigating quantum-radiation-pressure noise (QRPN) and the standard quantum limit (SQL) \cite{Purdy13}. When dominated by structural damping \cite{Saulson1990}, thermal noise rolls off with frequency faster than QRPN, so that at sufficiently high frequencies the SQL can be observed. However, thermal noise when dominated by viscous damping exhibits the same frequency dependence as QRPN, making it necessary to cool such a system (typically to below 1\,K) in order to observe the SQL.

Here we investigate thermal noise mechanisms in cantilever flexures, which are used in a wide range of opto-mechanical experiments \cite{Corbitt2007,Mowlowry2008,Purdy2013, Lubbe2013}, and feature in designs for mirror suspension systems in future gravitational wave detectors \cite{Ju2002, Punturo2010}. Typically, investigations of thermal noise measure bulk mechanical loss by exciting the system at one of its resonant frequencies and observing ringdown \cite{Ju2002, Duffy1992, Reid2006, Gretarsson2000}. However, this technique is limited to the resonant frequencies of the system, and in many high-precision devices (including current generation gravitational-wave detectors such as advanced LIGO \cite{LIGO2015}) it is the contribution of thermal noise \emph{away} from any mechanical resonance that is of most interest \cite{Black2004a, Numata2003}. This is considerably more difficult to measure due to the reduced amplitude of off-resonance fluctuations, and necessitates a more sensitive read-out technique as well as a highly isolated environment. To date, measurements have been reported showing coating and mirror thermal noise as the dominant source of fluctuations in various regions of the displacement spectrum \cite{Black2004b, Neben2012}.

Here we present direct measurements of broadband displacement spectra limited by suspension thermal noise across the audio-frequency band from 10\,Hz to 10\,kHz, which is particularly relevant to ground-based gravitational wave detectors. The recorded spectra span over a decade above and below the fundamental flexure resonances, mapping out the suspension thermal noise across these frequencies. Both a high-$Q$ material (niobium) and a low-$Q$ material (aluminium) are studied. We use cavity readout and Pound-Drever-Hall (PDH) locking \cite{Drever1983} to retrieve the flexure displacement. Our results are well explained by a simple model which includes both structural and thermoelastic damping. This agreement confirms the frequency dependence of these different thermal noise mechanisms, which is vital for experiments seeking to measure QRPN and the SQL, especially in the case of niobium which, with its low bulk loss, is a material of interest for such experiments.


\section{Experimental system}
\label{sect:system}

\begin{figure}[b]
\centerline{
\includegraphics[width=12cm]{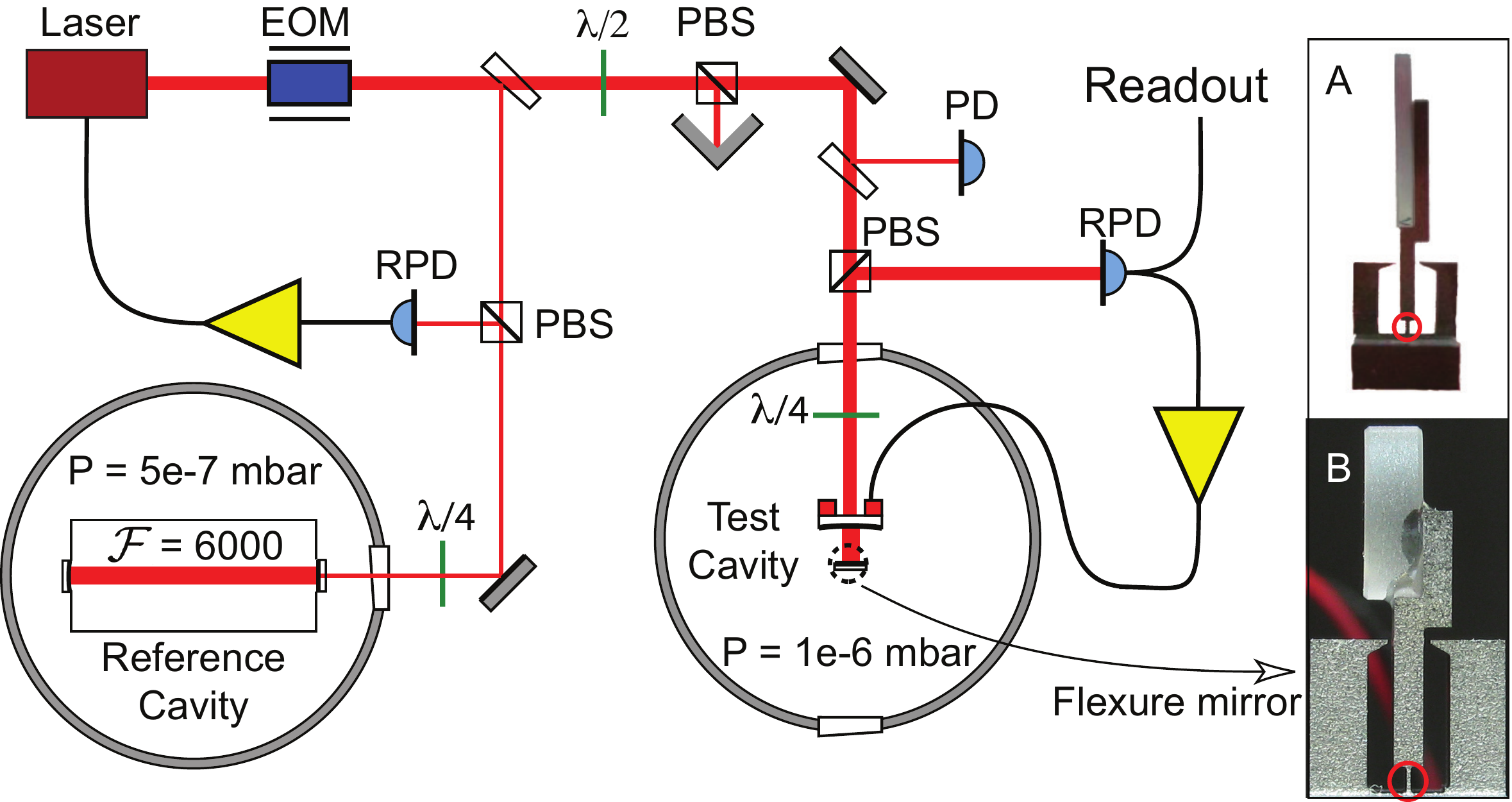}}
 \caption{Schematic of the experimental setup (PBS: polarising beam splitter, EOM: electro-optic modulator, PD: photodiode monitoring input power, RPD: photodiode measuring reflected power). The insets A and B show the rear mirror of the test cavity mounted on the niobium and aluminium flexures (red circles).}
 \label{fig:system}
\end{figure}

We designed monolithic inverted-pendulum flexures made from niobium and aluminium as shown in the inset of Figure \ref{fig:system}. The geometrical simplicity of the mechanical oscillators was intended to isolate the fundamental resonant frequencies from higher-order modes. The mechanical oscillators were manufactured by electric discharge machining.

The niobium flexure membrane was 6.35\,mm wide, 1\,mm high and 72\,$\mu$m thick. A dielectric mirror 7\,mm in diameter and 1\,mm thick was glued to the top of the flexure structure. The effective mass of the flexure (including the mirror) was 0.7\,g, resulting in a fundamental resonant frequency of 85\,Hz. The quality factor was independently determined from a ringdown measurement to be $Q = 44\,000$.

The aluminium flexure was more compact, with a membrane 5\,mm wide, 1\,mm high and 120\,$\mu$m thick. A mirror 1/4" in diameter and 2\,mm thick was glued to the top of the flexure. With an effective mass of 0.4\,g, the fundamental resonance is at 271\,Hz with $Q = 2\,200$, also independently determined using a ringdown measurement. The lower $Q$ material was chosen in order to increase the off-resonant thermal noise while reducing optically driven mechanical ringing.

To measure the extremely small thermal displacements with a high signal-to-noise ratio, the flexure is placed into a vacuum chamber and acts as the back mirror of a Fabry-Perot test cavity (TC). Displacements of the flexure imprint a phase shift onto the light bouncing off the mirror, which is amplified by approximately a factor of the cavity finesse $\mathcal{F} \sim 10^3$. The layout of the experiment is shown in Figure \ref{fig:system}. The TC was 12\,mm long and comprised a front mirror glued to a piezo-electric transducer (PZT) in addition to the back mirror on the flexure. The cavity finesse was $600$ and $700$ for the aluminium and niobium flexure experiments, respectively. The TC was kept on resonance using the Pound-Drever-Hall (PDH) locking technique \cite{Drever1983, Black2001}, and fluctuations in the cavity length were read out via the error signal of the PDH lock (labelled ``Readout'' in Figure \ref{fig:system}). 

To successfully measure thermal fluctuations using this technique, noise in the frequency of the interrogating laser must be reduced to below the equivalent thermal noise displacement, since these would otherwise appear as changes in the cavity length. To achieve this, the laser was locked to a high-finesse ($\mathcal{F} = 6000$), 20\,cm-long Zerodur reference cavity (RC) suspended in vacuum. It is also critically important that thermal fluctuations be the dominant source of changes in the cavity length. For this, the TC is suspended inside the vacuum chamber by a multi-stage vibration isolation system, which provides an effective `seismic wall' at 10\,Hz -- 40\,Hz.

The bandwidth of the PDH lock during the niobium flexure experiment was about 850\,Hz. As the flexure resonance was within the locking bandwidth, the PDH spectrum was corrected using the measured closed-loop servo response in order to obtain the thermal noise. In contrast, the servo unity gain frequency during the aluminium flexure experiment was less than 10\,Hz, well below the fundamental flexure resonance. The effect of the PDH lock at the flexure resonance was measured and found to be insignificant, thus no correction was required for these spectra. The aluminium flexure measurements were recorded using a digital control system, while those of the niobium flexure were recorded using a SR785 analog network analyser. All experiments were performed at room temperature (293\,K).


\section{Displacement noise measurements}
\label{sect:measurements}

\subsection{Aluminium flexure}

\begin{figure}[h]
\centerline{\includegraphics[width=13.5cm]{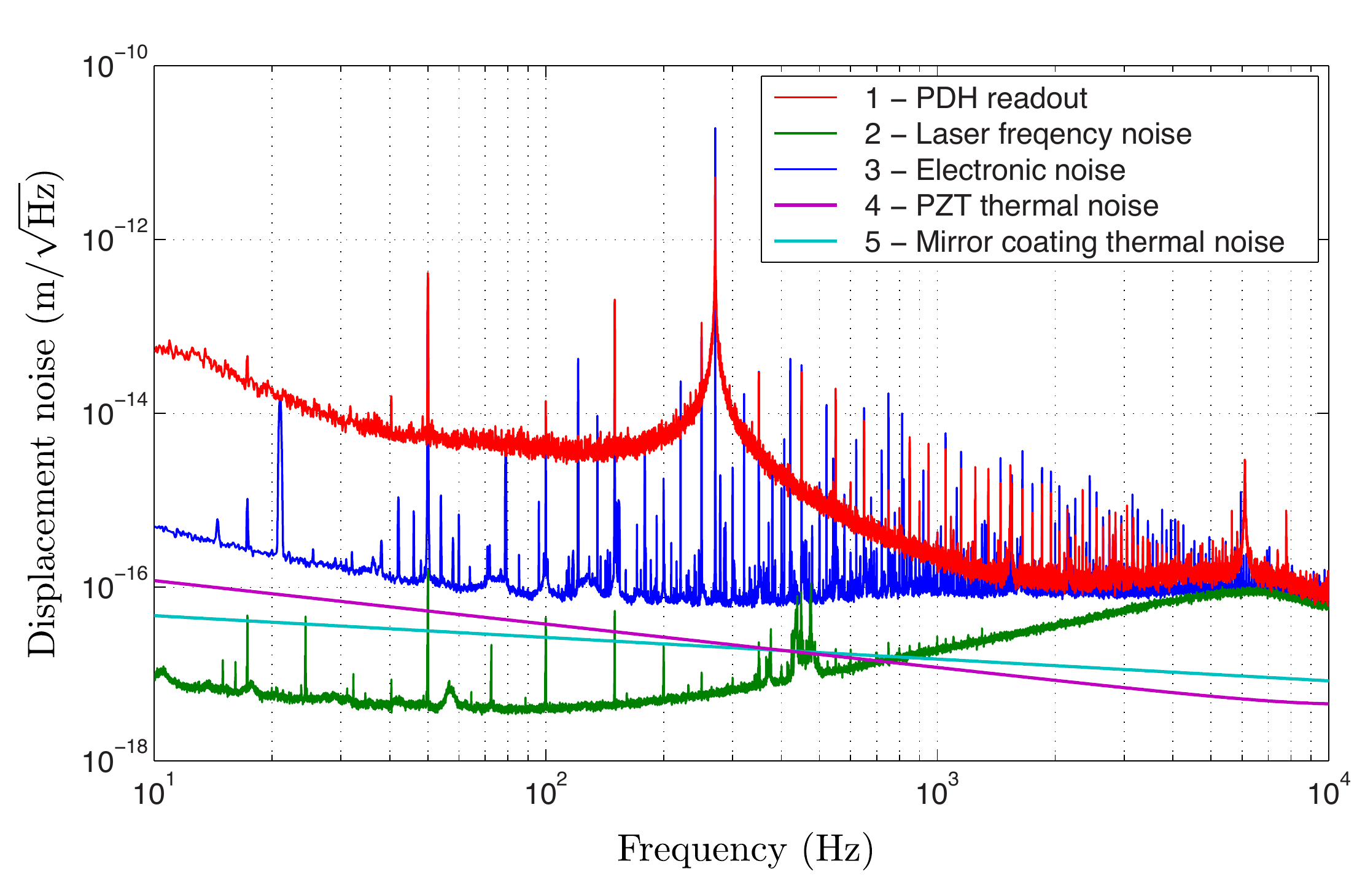}}
\caption{Thermal noise of the aluminium flexure. (1) PDH error signal readout; (2) gain-limited laser frequency noise, converted to an equivalent displacement noise; (3) electronic noise; (4) PZT thermal noise; (5) mirror coating thermal noise.}
\label{fig:measurementAl}
\end{figure}

The measured displacement fluctuation spectrum of the aluminium flexure is shown in Figure \ref{fig:measurementAl}. Also shown are laser frequency noise (converted to equivalent displacement noise), electronic noise and the expected thermal noise of the mirror coatings and PZT \cite{Braginsky2003}.\footnote{The PZT thermal noise is calculated using Equation (\ref{eqn:powerspectrum}) with the following parameters: $Q = 70$, $\omega_0 = 2\pi\times 40$\,kHz, $m$ = 11.5\,g.} The measured displacement spectrum is above all of these noise sources, and therefore expected to be dominated by thermal fluctuations, up to an order of magnitude above and below the fundamental mechanical resonance at 271\,Hz. Laser frequency noise begins to dominate around the second mechanical resonance at 6\,kHz. The displacement was calibrated providing an experimental uncertainty of 23\%.

\subsection{Niobium flexure}

\begin{figure}[ht]
\centerline{\includegraphics[width=13.5cm]{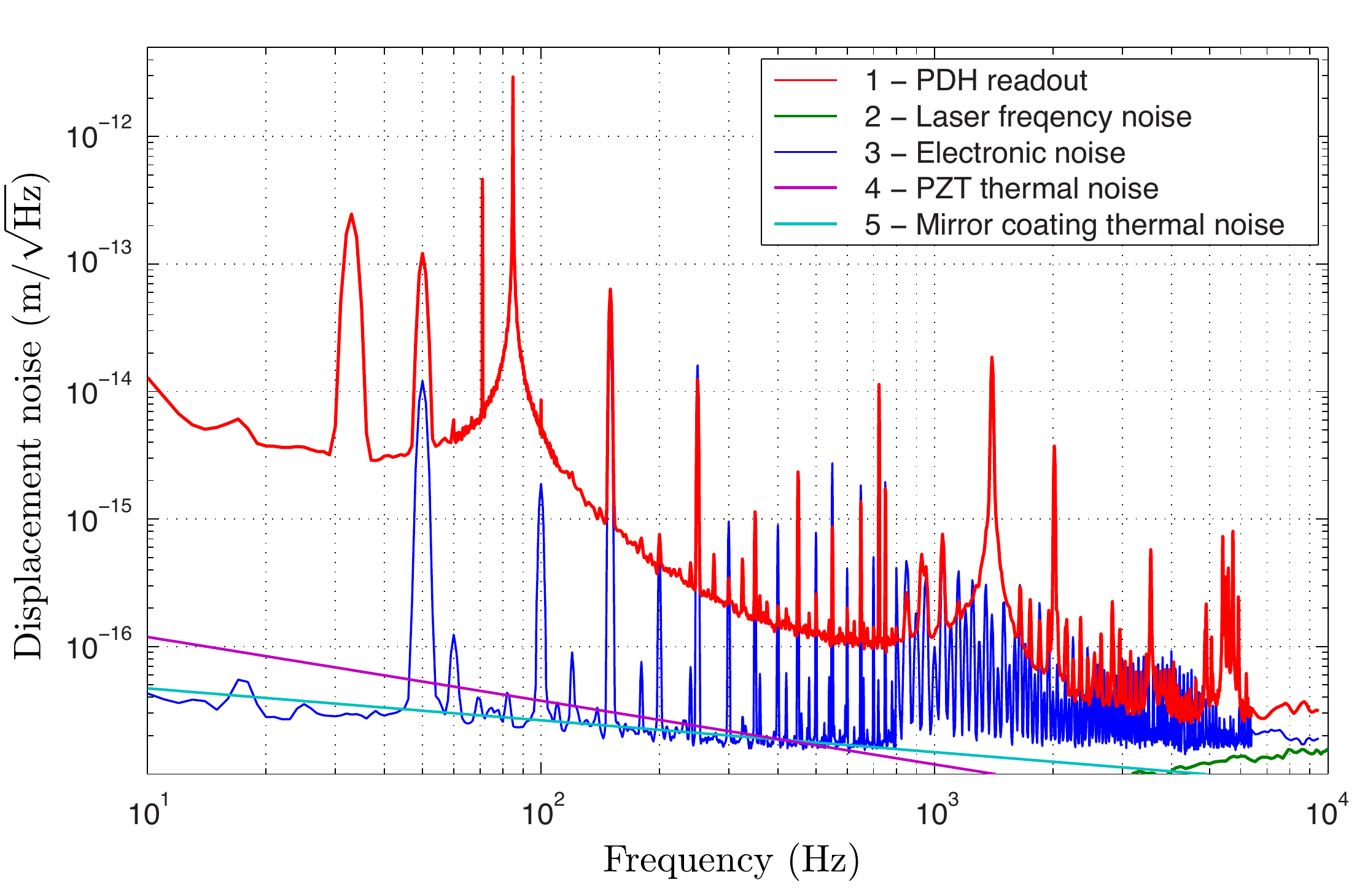}}
\caption{Thermal noise of the niobium flexure. (1) PDH error signal readout; (2) gain-limited laser frequency noise, converted to an equivalent displacement noise; (3) electronic noise; (4) PZT thermal noise; (5) mirror coating thermal noise.}
\label{fig:measurementNb}
\end{figure}

The measured displacement fluctuation spectrum of the niobium flexure is shown in Figure \ref{fig:measurementNb}, along with the laser frequency noise, electronic noise, and expected thermal noise of the PZT and mirror coatings. The measured spectrum is well clear of these other noise sources up to above the second mechanical resonance at around 1.4\,kHz. Calibration lines were injected at 33\,Hz and 723\,Hz, setting an experimental uncertainty on the displacement calibration of 17\%.


\section{Thermal noise model}
\label{sect:model}

Thermal noise is present in all macroscopic oscillating systems, driven by the thermal energy $k_B T$ present in every degree of freedom. From the fluctuation-dissipation theorem, the power spectrum of the thermal fluctuations $x_{\rm{th}}(t)$ in a harmonic oscillator at temperature $T$ can be determined from its mechanical response, characterized by oscillation frequencies $\omega_k$, their effective masses $m_k$, and corresponding losses $\phi_k(\omega)$, which are generally frequency-dependent \cite{Greene1952, Saulson1990}. The equation describing this power spectrum is:
\begin{equation}
\hat{X}_{\rm{th}}^{2}(\omega) = \sum_{k=0}^{n} { \frac{4k_B T\omega_k^2 \phi_k(\omega)}{m_k\omega \left[ \left(\omega_k^2 - \omega^2 \right)^2 + \omega_k^4\phi_k^2(\omega) \right] }} \,,
\label{eqn:powerspectrum}
\end{equation}
where $k_B$ is the Boltzmann constant. Here, the dimensionless loss parameter $\phi_k(\omega)$ represents the linear sum of all losses in the system for the $k^{\rm{th}}$ mode.

Due to the geometry of the flexures presented here, damping of all relevant mechanical modes is dominated by structural loss $\phi_{struc}$, due to internal friction, and viscous thermoelastic loss $\phi_{te,k}(\omega)$, caused by heat flow as different parts of the material are subjected to differential stresses. The total loss for a particular mode,
\begin{equation}
\phi_k (\omega) =  \phi_{struc}  + \phi_{te,k} (\omega) \,,
\label{eqn:totaldamping}
\end{equation}
can be obtained from ringdown measurements. The structural loss is the same for all modes and is independent of frequency \cite{Saulson1990}; in the absence of any other loss mechanisms, the inverse of the structural loss would determine the quality factor $Q$ of the oscillator. On the other hand, thermoelastic loss varies with frequency, and is dependent on the bulk material and geometry of the flexure. It can also be different for different oscillation modes. In general, the thermoelastic loss of a mode $k$ is described in terms of a strength $\Delta$ and characteristic time $\tau_k$ \cite{Cerdonio2001}, as:
\begin{equation}
\phi_{te,k} (\omega) = \Delta \frac{\omega \tau_k}{1 + (\omega \tau_k)^{2}} \,,
\label{eqn:tedamping}
\end{equation}
where
\begin{eqnarray}
\Delta = \frac{\alpha^{2} E_{y} T}{\rho C_{v}} \label{eqn:tedelta}  \,, \\
\tau_k = \frac{\rho C_{v} l_k^2}{\kappa \pi^2}  \label{eqn:tetau}  \,.
\end{eqnarray}
In these equations, $\alpha$, $E_y$, $\rho$, $C_v$ and $\kappa$ represent the linear thermal expansion coefficient, Young's modulus, density, specific heat and thermal conductivity of the flexure material, respectively. Values for these are given in Table \ref{table:parameters}. The parameter $l_k$ represents the path length along which heat flows as the material experiences stress and strain. This varies depending on the mode of oscillation; for example, the path length $l_0$ for the fundamental (bending) mode of our cantilever flexures is simply the membrane thickness, while for a higher-order shear mode $l_k$ could depend additionally on the width and height of the membrane. The characteristic time taken for heat to be transferred across this distance is given by $\tau_k$, and gives rise to a peak in the frequency response at $f_k = 1/\tau_k$.

\begin{table}[t]
\caption{Thermal noise model parameters.}
\centering
\begin{tabular}{| l | c | c | l |} \hline
Parameters 					& Aluminium 	   			& Niobium 				& Units \\\hline
Resonant frequency				& 271					& 85  					& Hz \\
Mirror diameter 				& $6.35  \times 10^{-3} $  		& $6.35  \times 10^{-3}$ 		& m  \\
Mirror thickness 				& $2  \times 10^{-3}$			&$1  \times 10^{-3}$	 		&m \\
Young's modulus ($E_y$)			& $71$					& $105$					& GPa	\\
Linear thermal expansion coefficient ($\alpha$)		 & $23 \times 10^{-6}$		& $7.3 \times 10^{-6}$		& $\mathrm{K}^{-1}$\\
Specific heat ($C_v$)			& $904$					& $265$					& $\mathrm{J(kgK)}^{-1}$		\\
Thermal conductivity	($\kappa$)	& $138$					& $54$					& $\mathrm{W(mK)}^{-1}$		\\
Density ($\rho$)				& $2820$					& $8578$					& $\mathrm{kgm}^{-3}$		\\
Thermoelastic resonance ($f_{te}$)	& $5.6 \times 10^{3}$		& $7.2 \times 10^{3}$		& Hz				\\\hline
\end{tabular}
\label{table:parameters}
\end{table}

\begin{figure}[b]
\centerline{
\includegraphics[width=15cm]{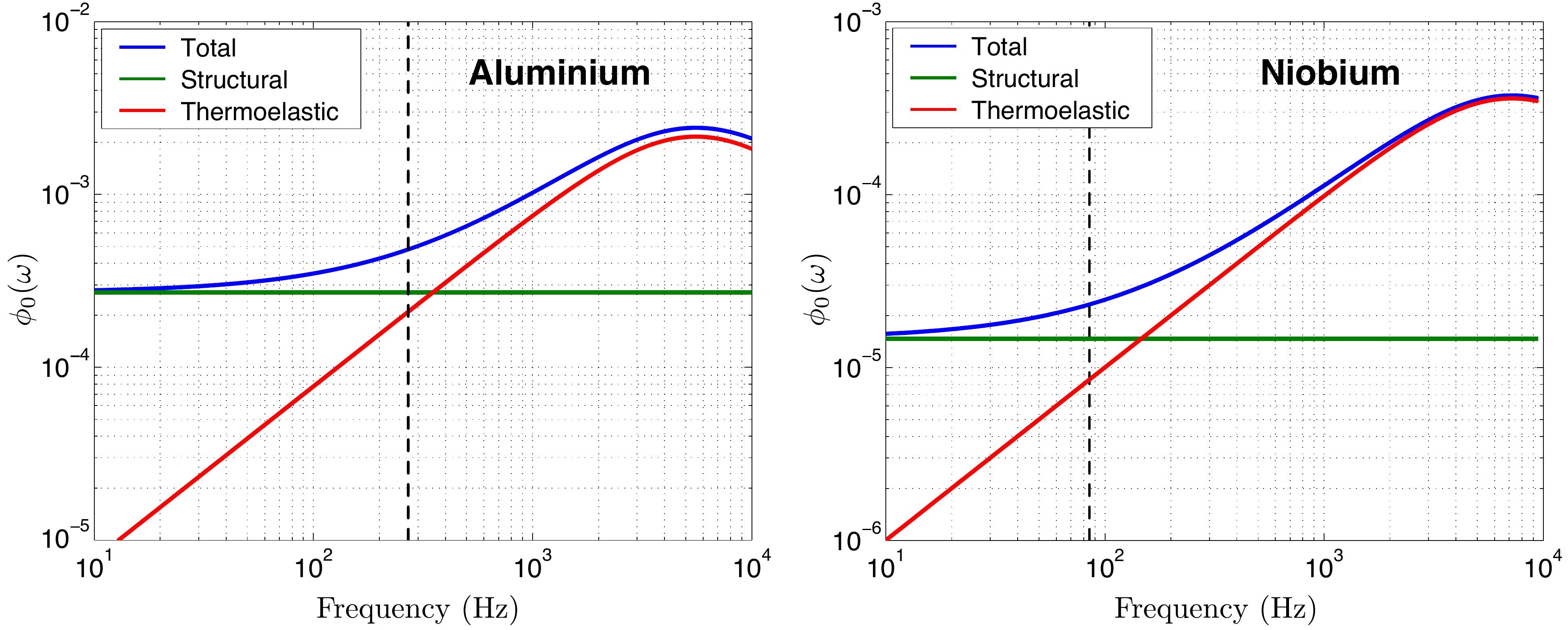}}
\caption{Calculated frequency-dependent loss $\phi_0(\omega)$ for the fundamental oscillation mode of the aluminium and niobium flexures, showing the contributions of structural $\phi_{struc}$ and thermoelastic $\phi_{te,0}(\omega)$ damping. The fundamental flexure resonance frequency for each material is indicated by the vertical dashed line. At the pressures used in the experiment, loss due to gas damping is insignificant and is thus omitted.}
\label{fig:theory}
\end{figure}

Figure \ref{fig:theory} shows the relative contribution of structural and thermoelastic processes to the total thermal loss for our flexures in the frequency range of interest. At the fundamental resonant frequencies (indicated by the vertical dashed lines), the thermoelastic losses $\phi_{te,0}(\omega_0)$ calculated from Equation (\ref{eqn:tedamping}) are $2.1\times10^{-4}$ (aluminium) and $8.5\times10^{-6}$ (niobium). The total loss was obtained from a ringdown measurement of $Q$, allowing us to deduce from Equation (\ref{eqn:totaldamping}) a structural loss of $2.7\times10^{-4}$ for aluminium and $1.5\times10^{-5}$ for niobium. The thermoelastic loss peak occurs at around 5.6\,kHz and 7.2\,kHz for aluminium and niobium, respectively.

In both cases, structural loss is the dominant noise source at frequencies below the fundamental resonance, while at higher frequencies the loss is dominated by thermoelastic damping. This feature is particularly noteworthy for QRPN experiments due to the different frequency dependence of these loss mechanisms \cite{Purdy13}. Thermal noise originating from structural loss exhibits a $1/f^{2.5}$ rolloff, so that radiation pressure noise, which rolls off as $1/f^2$, can dominate at high frequencies. On the other hand, fluctuations due to thermoelastic loss have the same $1/f^2$ frequency dependence as radiation pressure, and can therefore mask QRPN unless the system is cooled to cryogenic temperatures. In such experiments, a model such as the one used here can be used to ensure that QRPN will dominate the fluctuation spectrum by adjusting the flexure geometry to shift the thermoelastic peak out of the frequency band of interest.


Other loss mechanisms were also investigated, including damping from residual background gas collisions, loss from the flexure clamping and the amount and type of glue used. None of these tests showed significant changes to the off-resonant thermal noise.

\section{Comparison}

\begin{figure}[b]
\centerline{\includegraphics[width=16cm]{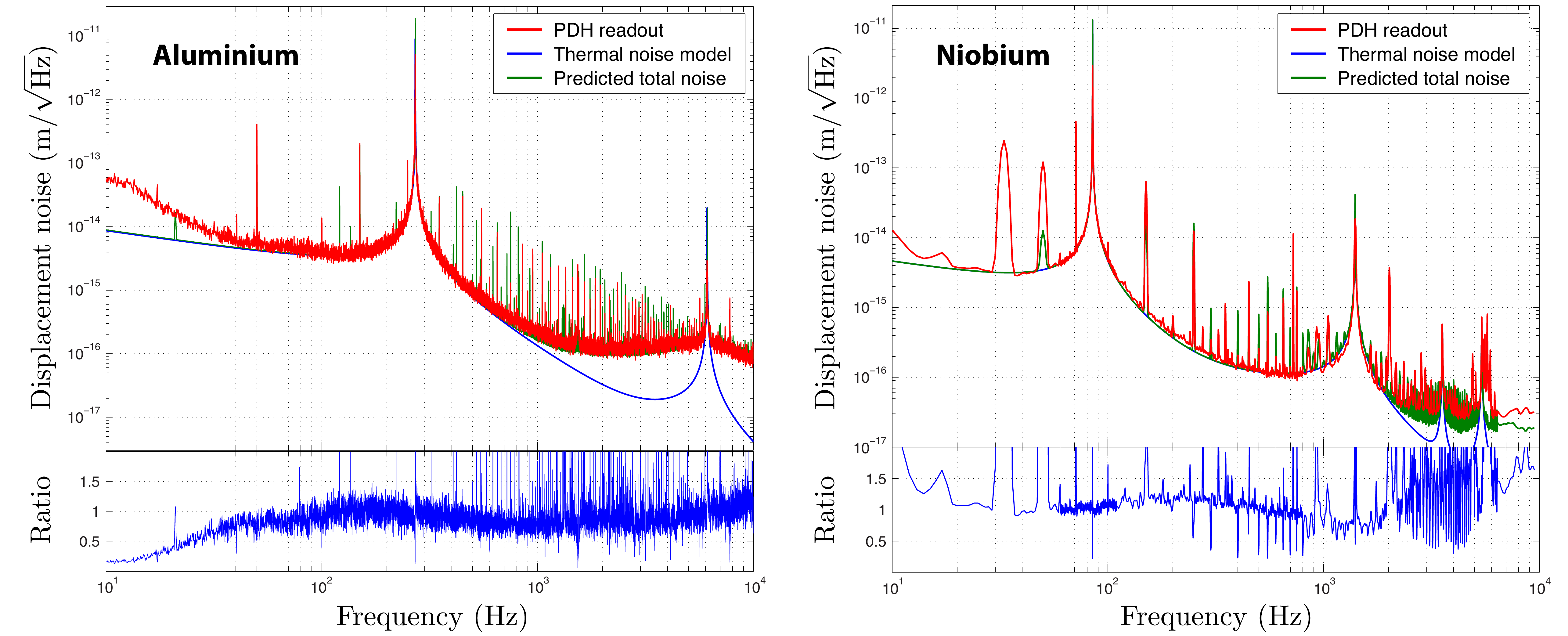}}
\caption{Comparison of thermal noise measurement and model predictions for aluminium and niobium flexures. The traces show the PDH error signal readout (red), the sum of structural and thermoelastic noise as predicted by the model detailed in Section \ref{sect:model} (blue), and the quadrature sum of the thermal noise model and other experimental noise contributions (green). The lower plots show the ratio between the measured PDH error signal and the predicted total noise, which in both cases is close to unity between 50\,Hz and 5\,kHz.}
\label{fig:comparison}
\end{figure}

We now compare the predictions of the model developed above with our measured thermal noise displacement spectra.

Figure \ref{fig:comparison} shows the measured displacement noise spectra of the flexures overlaid with the thermal noise model developed above (Equation \ref{eqn:tedamping}) added in quadrature with the experimental noise sources discussed in Section \ref{sect:measurements}. No fitting was performed, and the ratio of the experimental and theoretical traces is also given for the purposes of comparison. For the aluminium flexure, the measured displacement deviates from the predicted total noise below 50\,Hz, which is due to residual seismic coupling into the final test cavity suspension stage and spurious scattering of light onto the reflection photodiode. The niobium measurement follows the predicted noise trace well from 10\,Hz up to 2\,kHz.


\section{Conclusions}

We have reported measurements of off-resonance thermal noise for aluminium and niobium cantilever flexures in the audio frequency band between 10\,Hz and 10\,kHz, using cavity readout and Pound-Drever-Hall (PDH) locking to observe the displacement of the flexures due to thermal fluctuations at room temperature. Our results are of particular interest to the design of suspension systems for next-generation gravitational wave detectors. Our experimental results show good agreement with a simple model which includes structural damping and frequency-dependent thermoelastic damping. The ability of such a model to predict thermal noise behavior has implications for the design of experiments on quantum-radiation-pressure noise and the standard quantum limit.

\ack
We acknowledge Dr.\ Li Ju for her help with manufacturing the niobium flexure. This work was supported by the Australian Research Council.

This publication has the LIGO document number LIGO-P1500089.

\section*{References}

\bibliographystyle{unsrt}
\bibliography{references}

\end{document}